\documentclass[12pt]{article}

\usepackage{scicite}
\usepackage[pdftex]{graphicx}

\usepackage{times}

\newenvironment{sciabstract}{%
\begin{quote} \bf}
{\end{quote}}

\newcounter{lastnote}
\newenvironment{scilastnote}{%
\setcounter{lastnote}{\value{enumiv}}%
\addtocounter{lastnote}{+1}%
\begin{list}%
{\arabic{lastnote}.}
{\setlength{\leftmargin}{.22in}}
{\setlength{\labelsep}{.5em}}}
{\end{list}}

\title{Radio Frequency Association of Efimov Trimers} 

\author
{Thomas Lompe,$^{1,2,3\ast}$ Timo B. Ottenstein,$^{1,2,3}$ Friedhelm Serwane,$^{1,2,3}$\\
Andre N. Wenz,$^{1,2}$ Gerhard Z\"urn$^{1,2}$ and Selim Jochim$^{1,2,3}$ \\
\\
\normalsize{$^{1}$Physikalisches Institut, Ruprecht-Karls-Universit\"at Heidelberg, Germany}\\
\normalsize{$^{2}$Max-Planck-Institut f\"ur Kernphysik, Saupfercheckweg 1, 69117 Heidelberg, Germany}\\
\normalsize{$^{3}$ExtreMe Matter Institute EMMI, GSI Helmholtzzentrum f\"ur Schwerionenforschung, Darmstadt, Germany}\\
\\
\normalsize{$^\ast$To whom correspondence should be addressed; E-mail:  thomas.lompe@mpi-hd.mpg.de.}
}

\date{}

\begin{document} 




\maketitle


\begin{sciabstract}
  The quantum-mechanical three-body problem is one of the fundamental challenges of few-body physics. 
When the two-body interactions become resonant, an infinite series of universal three-body bound states is predicted to occur, whose properties are determined by the strength of the two-body interactions. 
We report on the association and direct observation of a trimer state consisting of three distinguishable fermions using radio-frequency (RF) spectroscopy.
The measurements of its binding energy are consistent with theoretical predictions which include non-universal corrections.

\end{sciabstract}

Under certain conditions the long-range behavior of a physical system can be described without detailed knowledge of its short-range properties; a prime example of this concept of universality are few-body systems with resonant interactions\cite{braaten_review}.
Ultracold gases, where resonant scattering may be achieved by tuning the interactions using Feshbach resonances \cite{Feshbach_review}, have been used extensively to test the predictions of universal theory.

If the parameter describing the interactions, the s-wave scattering length $a$, is much larger than the characteristic length scale $r_0$ of the interaction potential the few-body physics in such ultracold gases is predicted to become universal. 
For two particles with a large positive scattering length there is a weakly bound universal dimer state whose binding energy scales as $1/a^2$.
 For three particles with large interparticle interactions there is a series of weakly bound trimer states, which are called Efimov trimers\cite{efimov}. For negative scattering lengths, these trimer states
become bound at critical values of the interaction strength, which are spaced
by a universal scaling factor. For diverging interaction strength this results in an infinite series of Efimov states.
Their absolute position depends on short-range three-body physics, which can be described by a single three-body parameter.
For positive scattering lengths, the trimer states disappear when they cross the atom-dimer threshold, i.e. their binding energy becomes degenerate with the binding energy of a dimer state (Fig. 5).
These crossings of trimer states with the continuum threshold can be observed experimentally as resonant enhancements of the rate constants for inelastic three-atom and atom-dimer collisions, respectively. This allowed to obtain the first convincing evidence for the existence of Efimov trimers by tuning the interparticle interaction in an ultracold atomic gas using a Feshbach resonance, which allowed to observe signatures of Efimov trimers in the rate of inelastic three-body collisions\cite{innsbruck_efimov}. Since then, this technique has been used with great success, but it is limited to observations of the crossings of Efimov states with the continuum \cite{innsbruck_atomdimer,wir_unten,ohara_unten,lens_hetero,lens_homo,ohara_oben,lev_li7,randy_li7,lev_li7_new,wir_atomdimer,naidon_atomdimer,greene_4body,innsbruck_4body}.

The predictions of universal theory can be tested by observing multiple crossings of trimer states with the continuum in a single system. Some experiments are consistent with these predictions\cite{lev_li7,lev_li7_new}, whereas others see larger deviations\cite{lens_homo} or even a systematic shift across a Feshbach resonance\cite{randy_li7}. We study Efimov physics in a conceptually different manner by directly measuring the binding energy of an Efimov state as a function of interaction strength. 

We use an ultracold Fermi gas consisting of fermionic $^6$Li atoms in the three energetically lowest Zeeman substates.
We label these states $\left|1\right\rangle$, $\left|2\right\rangle$ and $\left|3\right\rangle$\cite{wir_unten}. Because of Pauli blocking, only atoms in different states interact with each other at ultracold temperatures.  These interactions are described by three different interparticle scattering lengths $a_{12}$, $a_{23}$ and $a_{13}$ for the respective combinations of atoms in different states\cite{bartenstein_julienne} (Fig. 1 A). For each of these combinations there is a broad Feshbach resonance, leading to three different weakly bound dimer states, which we label $\left|12\right\rangle$, $\left|23\right\rangle$ and $\left|13\right\rangle$ (Fig. 1 B). If the corresponding scattering length is much larger than the characteristic radius $r_0 \approx 60\,a_0$ of the $^6$Li interaction potential the dimer binding energy $E_{ij}$ is given by the universal relation $E_{ij} = \hbar^2/ m a_{ij}^2$, where $m$ is the mass of a $^6$Li atom, $\hbar = h/2\pi$ is the reduced Planck constant and the indices $i,j = 1,2,3$ indicate the state of the atoms.  
As the Feshbach resonances for the three combinations overlap, all scattering lengths can be tuned to large values simultaneously, leading to the appearance of Efimov states.
However, as the different scattering lengths do not diverge at the same magnetic field the series of trimer states is finite. There are two Efimov trimer states in this system\cite{braaten_6li} whose crossings with the continuum have all been located by observing enhanced inelastic collision rates\cite{wir_unten, ohara_unten, ohara_oben, wir_atomdimer, naidon_atomdimer} (Fig. 1 B). 
From these measurements the binding energies of these trimer states have been calculated using universal theory\cite{braaten_6li}.

\begin{figure}
\centering
\includegraphics [width= 8cm] {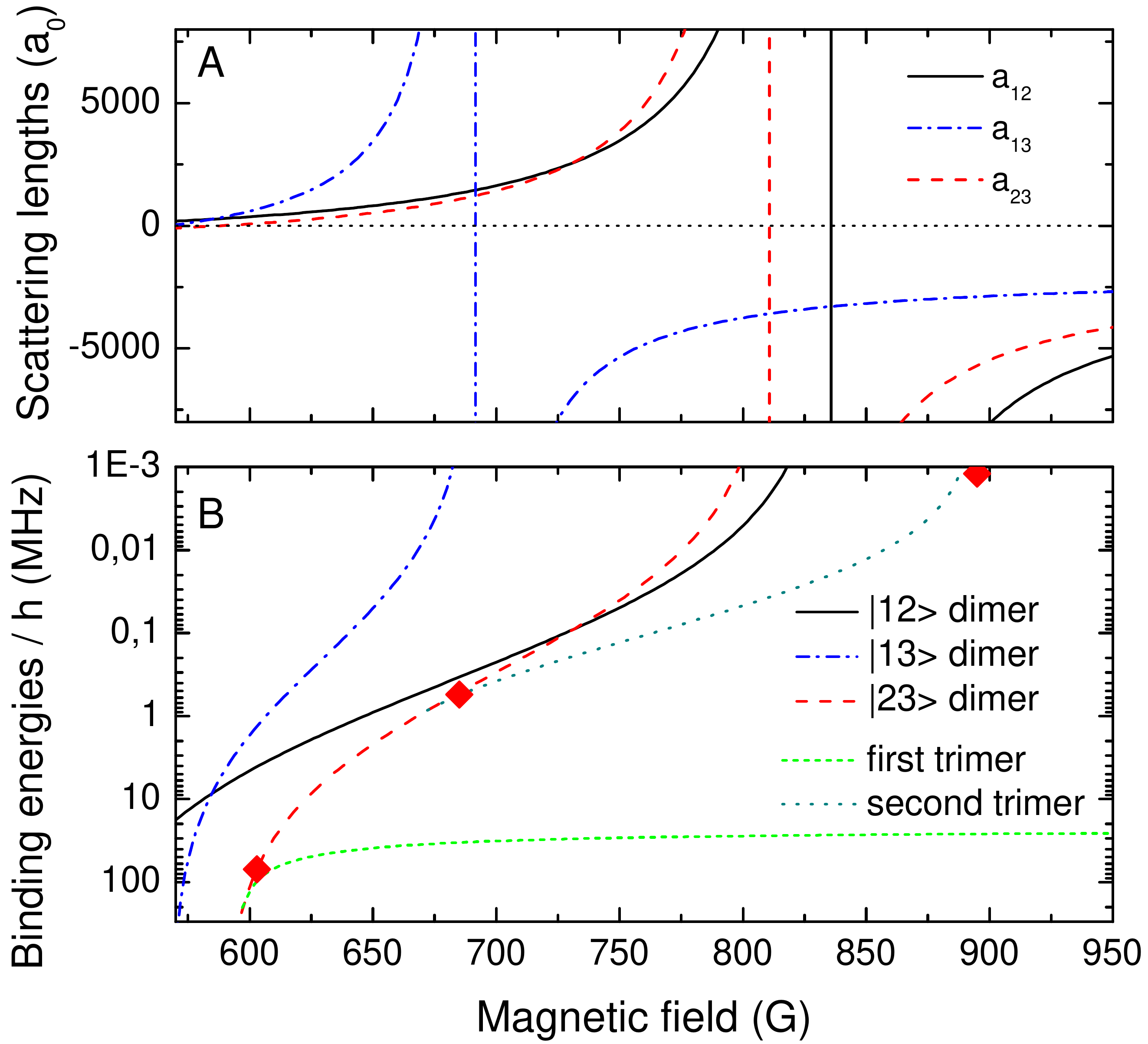}
\caption{(A) Values of the three different two-body scattering lengths for magnetic fields between $570$\,G and $950$\,G given in units of Bohr's radius $a_0$\cite{bartenstein_julienne}. (B) Binding energy of the dimer and trimer states according to universal theory, calculated using the three-body parameter determined from the crossing at $B = 895$\,G\cite{braaten_6li}. The measured positions of the crossings of the trimer states with the continuum are marked by the red diamonds\cite{ohara_oben, wir_atomdimer,naidon_atomdimer}.}
\label{fig:a_and_eb}
\end{figure}

To directly measure the binding energy of one of these Efimov states we use RF-spectroscopy, a technique which has already been used extensively to study weakly bound dimer states\cite{jin_RF, bartenstein_julienne}. This technique is based on driving transitions between different internal states of the atoms 
using RF-fields, where in both the initial and final state the atoms can either be free or bound in a molecule. The difference in binding energy between the initial and final state can be measured as a shift of the transition frequency from the bare atomic transition. 

The most straightforward way to do RF-spectroscopy of a bound state is to dissociate the molecule into free atoms. In this case the RF-transition is simply shifted by the binding energy of the molecule. However, as Efimov trimers are highly unstable with lifetimes ranging from a few ns to tens of $\rm{\mu}$s\cite{braaten_6li} it is impossible to prepare macroscopic samples of trimers in current experiments, and dissociation spectroscopy is not technically feasible.
This can be overcome by using RF-fields to associate trimers from free atoms or atoms and dimers, but obtaining observable association rates is challenging for several reasons.

Let us first consider the initial states available for the association.
Starting from one atom in state $\left|1\right\rangle$ and two atoms in state $\left|2\right\rangle$ one can use an RF-transition 
to drive the atoms in state $\left|2\right\rangle$ to state $\left|3\right\rangle$. However, this is a coherent process which affects both atoms in state $\left|2\right\rangle$ in the same way, so they remain identical particles. Thus, this process cannot lead to the formation of trimers. Instead we can bind two of the atoms into a weakly bound $\left|12\right\rangle$ dimer, which breaks the symmetry between the two atoms in state $\left|2\right\rangle$ as for one of them the RF-transition is now shifted by the binding energy of the dimer.
Then one has to consider the wavefunction overlap of the initial and final state.
Close to the crossing of the trimer state with the atom-dimer threshold the size of the dimer and the trimer are on the order of the scattering lengths. Hence, the spatial wavefunction of a dimer and an atom has a finite overlap with the spatial wavefunction of the trimer when the atom approaches the dimer to a distance on the order of the scattering length. Therefore, the association rate depends critically on the phase-space density of the initial atom-dimer mixture. If we were to use a degenerate gas it would spatially separate into a molecular Bose-Einstein-Condensate of dimers and an outer shell of unbound fermionic atoms, which would reduce the spatial overlap between atoms and dimers in the trap. Therefore we have to use a thermal mixture of atoms in state $\left|2\right\rangle$ and $\left|12\right\rangle$ dimers, which limits the efficiency of the association. A similar limitation has already been observed in the RF-association of weakly bound dimers\cite{arlt_association}. 
Finally, the short lifetime of the Efimov trimer leads to a suppression of trimer formation analogous to the quantum-Zeno effect. This can in principle be avoided by driving the RF-transition with a Rabi frequency which is large compared to the decay rate of the trimer. However, this is not feasible with our current experimental setup, which further reduces the association rate.

\begin{figure}
\centering
\includegraphics [width= 8cm] {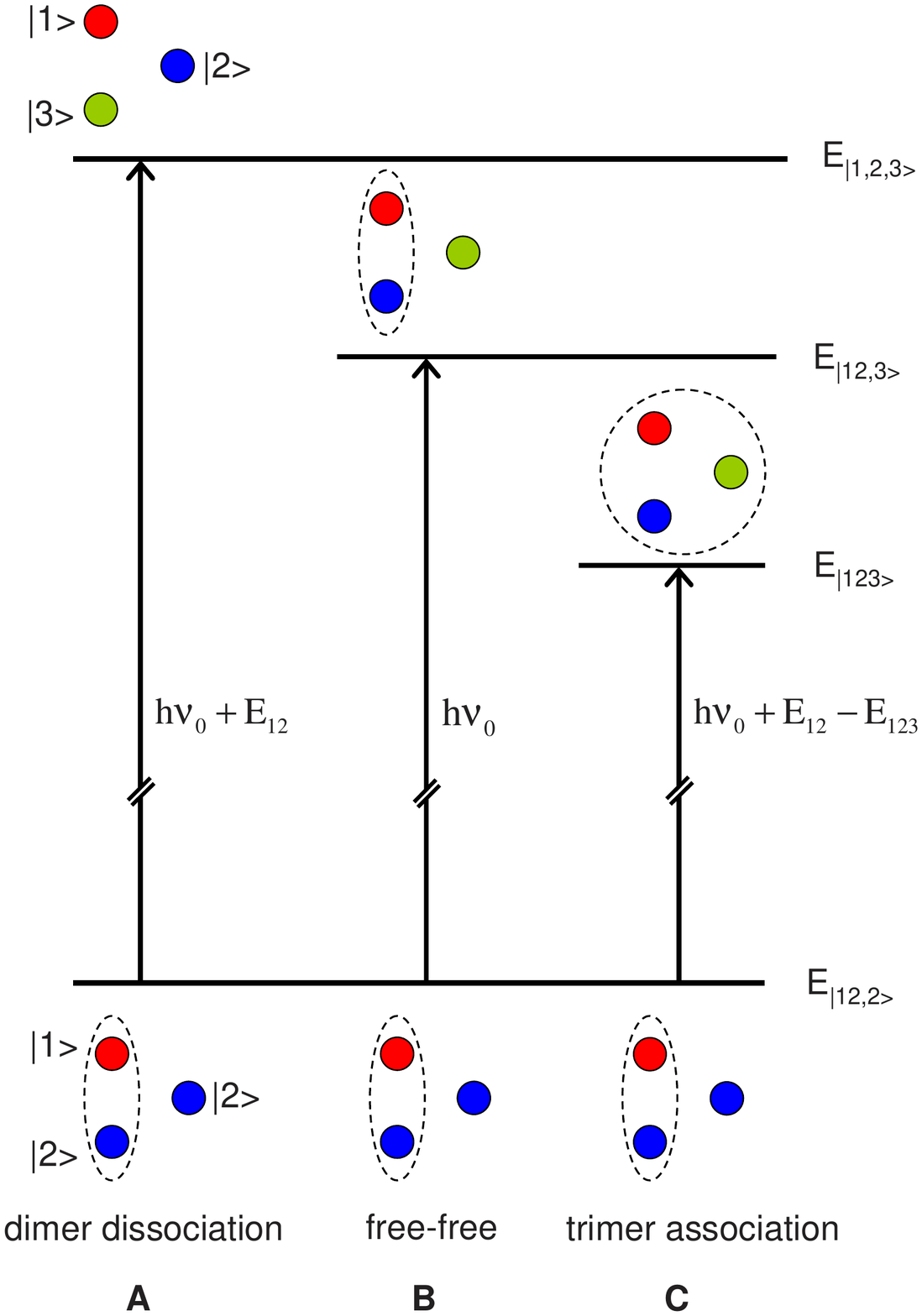}
\caption{Relevant transitions and energy levels for the RF-association of trimers. The initial system has the energy $E_{\left|12,2\right\rangle}$ of a free atom in state $\left|2\right\rangle$ and a $\left|12\right\rangle$ dimer. The frequency $\nu_0$ of the bare transition drives the unbound atoms from state $\left|2\right\rangle$ to state $\left|3\right\rangle$, leading to a $\left|3\right\rangle$-$\left|12\right\rangle$ atom-dimer mixture (B). By choosing the appropriate detuning from the bare atomic transition we can either dissociate the dimer, leading to three unbound atoms in states $\left|1\right\rangle$, $\left|2\right\rangle$ and $\left|3\right\rangle$ (A) or associate a dimer and an atom into a trimer (C). }
\label{fig:rf_scheme}
\end{figure}

From these considerations follows that the trimer state with the most favorable conditions for RF-association in our system is the second trimer state, as it is larger and has a longer lifetime\cite{braaten_6li}. This state crosses into the  $\left|1\right\rangle$-$\left|23\right\rangle$ atom-dimer continuum at $B = 685$\,G (Fig. 1 B).
To measure its binding energy, we prepare mixtures of atoms in state $\left|2\right\rangle$ and $\left|12\right\rangle$ dimers at magnetic fields between $670$\,G and 740\,G and apply RF-fields at frequencies around the $\left|2\right\rangle$-$\left|3\right\rangle$ transition. 

At the frequency $\nu_0$ which corresponds to the bare atomic transition, the free atoms are driven from state $\left|2\right\rangle$ to state $\left|3\right\rangle$, while the dimer remains unaffected (Fig. 2 B). If the RF is blue-detuned by the binding energy $E_{12}$ of the dimers, the dimers are dissociated (Fig. 2 A). The trimer is associated at the frequency $\nu = \nu_0 - E_{123}/h + E_{12}/h$, where $E_{123}$ is the binding energy of the Efimov trimer with respect to the $\left|1\right\rangle$-$\left|2\right\rangle$-$\left|3\right\rangle$ continuum.
Therefore, the association is red-shifted from the bare transition by the difference between the dimer and trimer binding energies (Fig. 2 C).
In each of these cases we observe a loss of atoms from the trap, either through decay in inelastic $\left|3\right\rangle$-$\left|12\right\rangle$ or $\left|1\right\rangle$-$\left|2\right\rangle$-$\left|3\right\rangle$ collisions  or through the decay of the associated trimers. If the RF is not resonant for either of these processes the atom number is not affected. Sample spectra for different magnetic fields are shown in Fig. 3.

\begin{figure}
\centering
\includegraphics [height= 15cm] {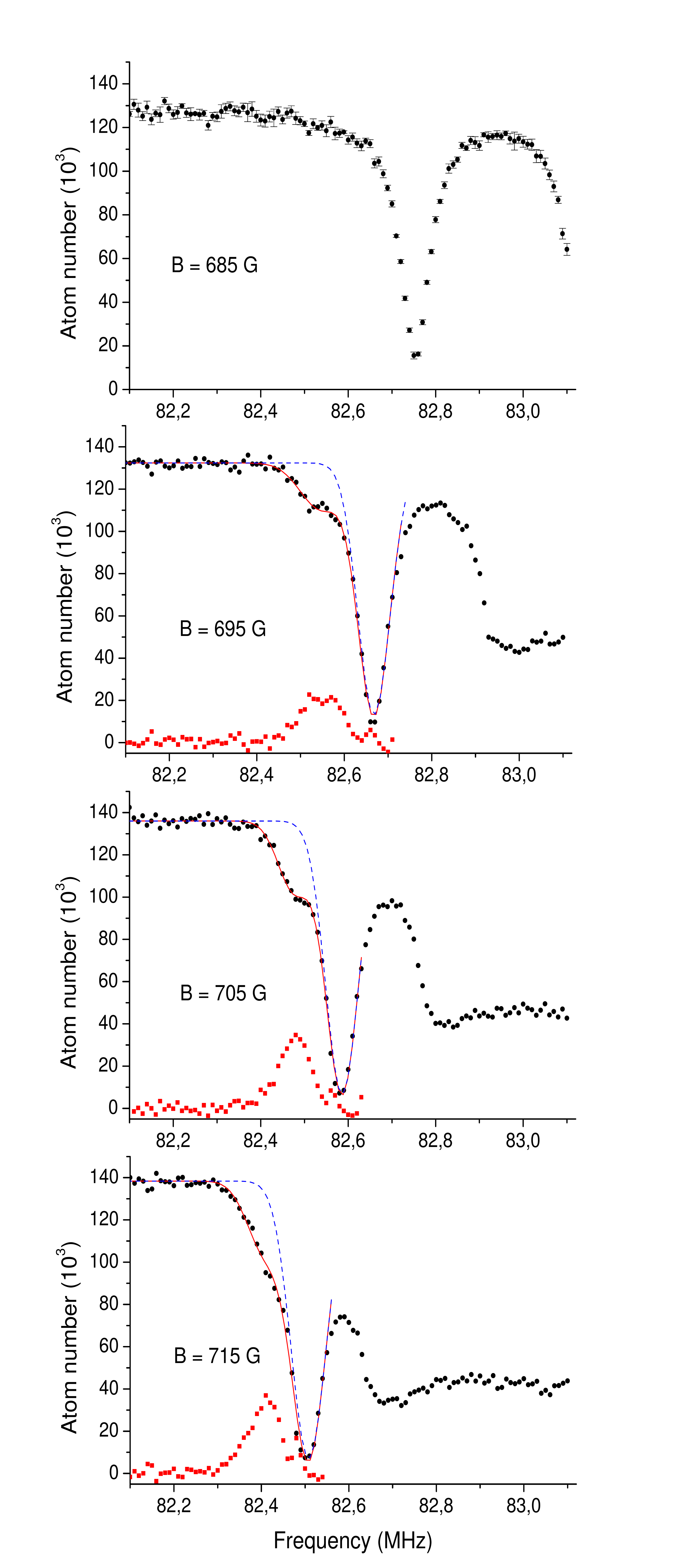}
\caption{Number of atoms in state $\left|2\right\rangle$ remaining after the RF-pulse as a function of frequency for four different magnetic fields (black circles). Each data point is the average of five to eight individual measurements, to improve clarity error bars are only shown for the dataset at 685\,G. The central dip results from driving the bare atomic transition and the dip to the right is a consequence of the dissociation of dimers. The trimer association manifests itself as a smaller feature sitting on the left flank of the central dip. The red lines are a fit of two overlapping Gaussians to the data. The dashed blue line shows the Gaussian fitted to the central dip. The difference between this fit and the data, which shows the signal from the trimer association, is plotted as red squares.}
\label{fig:sample_plots_12_2}
\end{figure}

Because of the limited wavefunction overlap of the initial and final state and the quantum-Zeno suppression of the association, we apply strong RF-pulses with a duration of 35-50 ms and a Rabi frequency of $\Omega \approx 2 \pi \times 7$\,kHz to associate enough trimers to obtain an observable decrease in the atom number. Collisions lead to decoherence and thus a strong broadening of the bare transition. 
However, as the association is offset from the free-free transition by $(E_{123} - E_{12})/h$, the association features can still be observed (Fig. 3). For magnetic fields below the crossing of the trimer and the $\left|1\right\rangle$-$\left|23\right\rangle$ atom-dimer threshold at $685$\,G there is no trimer state and consequently we observe no association peak. At $B=695$\,G the association signal is clearly visible. For higher magnetic fields $E_{123} - E_{12}$ becomes smaller, and the association peak moves closer to the free-free transition.
As we do not have a better model for the lineshapes of the broadened transitions we fit the spectra with two overlapping Gaussians to determine the position of the association dips. From these we calculate the binding energy of the trimer using the values of the binding energy of the $\left|12\right\rangle$-dimer known from previous experiments\cite{bartenstein_julienne}. The resulting binding energies of the trimer are plotted as black squares in Fig. 4. To confirm that the results are independent of the initial system and the transition used we performed the same measurement starting from a $\left|2\right\rangle$-$\left|23\right\rangle$ mixture at magnetic fields between $705$\,G and $725$\,G and driving the $\left|1\right\rangle$-$\left|2\right\rangle$ transition (Fig. 6). The two sets of measurements agree within experimental uncertainties (Fig. 4).

\begin{figure}
\centering
\includegraphics [width= 8cm] {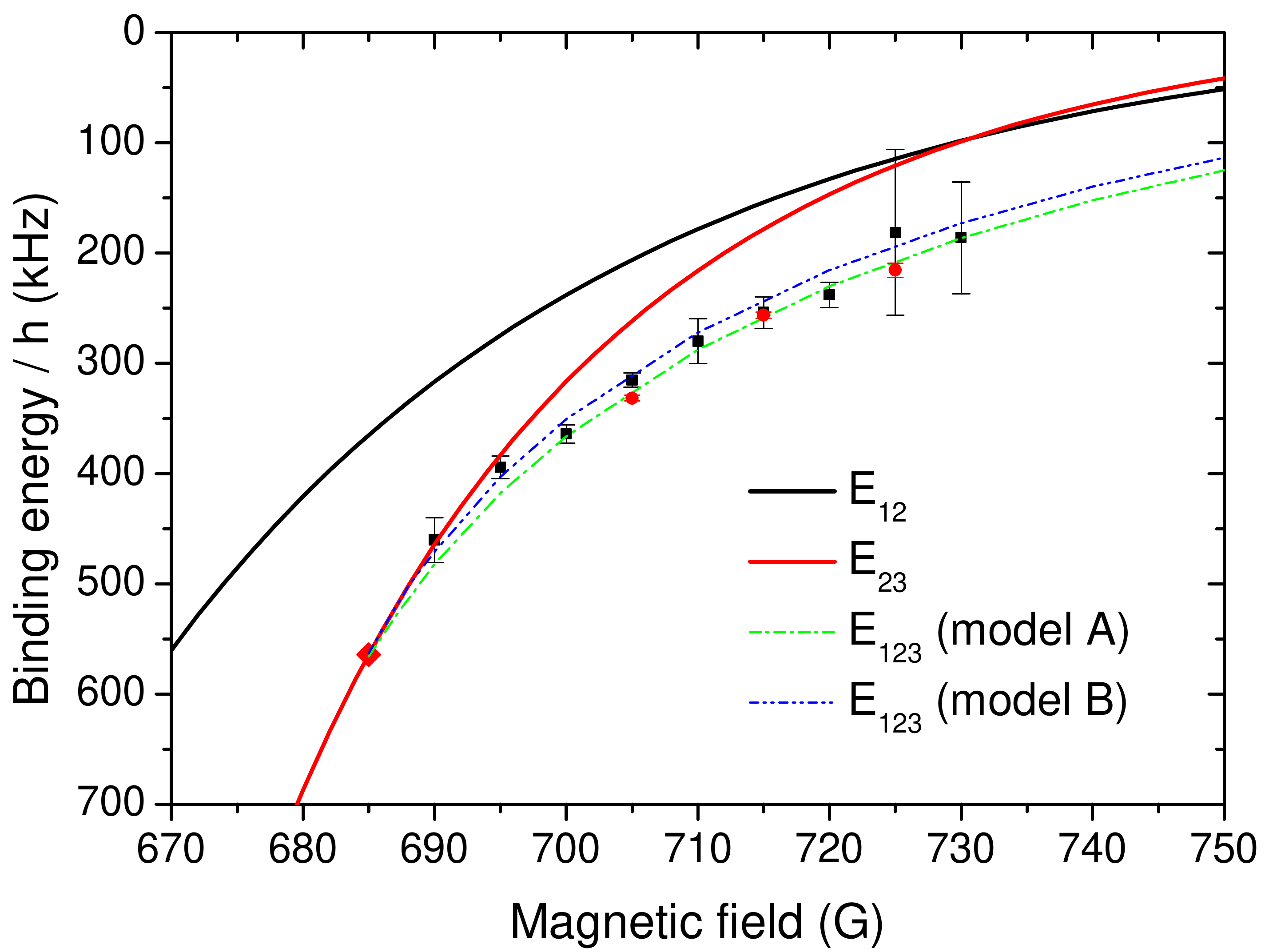}
\caption{Measured binding energy of the second Efimov trimer in $^6$Li measured using a $\left|2\right\rangle$-$\left|12\right\rangle$ (black squares) or $\left|2\right\rangle$-$\left|23\right\rangle$ (red dots) mixture as the initial system. The error bars are derived from the 2$\sigma$ confidence bounds of the fit to the RF-spectra. Additionally, there are systematic uncertainties caused by the broadened lineshapes and temperature effects (see methods section). The solid lines give the binding energies of the $\left|12\right\rangle$ and $\left|23\right\rangle$ dimers including non-universal corrections\cite{naidon_atomdimer}. The red diamond marks the position of the observed crossing of the trimer state with the $\left|1\right\rangle$-$\left|23\right\rangle$ atom-dimer threshold\cite{wir_atomdimer, naidon_atomdimer}. The dashed lines show the binding energy of the trimer according to calculations from\cite{naidon_atomdimer}.}
\label{fig:measured_eb}
\end{figure}

The measured binding energy of the trimer increases with decreasing scattering lengths while the difference to the binding energy of the $\left|23\right\rangle$-dimer decreases, as expected from the universal scenario. At $B=690$\,G the dimer and trimer binding energies are the same within our experimental resolution, and for $B \leq 685$\,G the trimer association feature vanishes.
This coincides with the previously observed enhancement of inelastic atom-dimer collisions at $685 \pm 2$\,G\cite{wir_atomdimer}, which confirms the interpretation of such loss resonances as signatures of a crossing of a trimer state with the continuum.
Additionally, we compare our data to the predictions from a theoretical model by Nakajima et al. which uses the observed crossings of the trimer states as inputs to calculate the trimer binding energies\cite{naidon_atomdimer}. This model includes finite range corrections to the two-body interactions known from previous experiments\cite{bartenstein_julienne}. 
To exclude a dependence of their results on the specific model, Nakajima et al. use two different parametrizations of the non-universal corrections (A and B) which yield essentially identical results. 
Within the errors our data is in good agreement with their predictions. 
Additionally, their model includes an energy dependence of the three-body parameter. However, in the magnetic field region of interest the difference to a model with a constant three-body parameter derived from the atom-dimer resonance at 685\,G is too small to be resolved with our current resolution\cite{naidon_atomdimer,pascal_french}.
By increasing the resolution of the measurements and reducing the systematic uncertainties it should be possible to use the techniques developed in this work to do precision tests of few-body physics and to directly determine the lifetime of Efimov trimers from the width of the association peaks. Driving the RF-transition with a Rabi frequency larger than the decay rate of the trimers should allow to coherently populate the trimer state as it was done for universal dimers\cite{arlt_association}. This would be a major step toward the preparation of macroscopic samples of Efimov trimers.

\bibliography{trimer}

\begin{thebibliography}{10}

\bibitem{braaten_review}
E.~Braaten, H.~W. Hammer, {\it Physics Reports\/} {\bf 428}, 259 (2006).

\bibitem{Feshbach_review}
C.~Chin, R.~Grimm, P.~Julienne, E.~Tiesinga, {\it Rev. Mod. Phys.\/} {\bf 82},
  1225 (2010).

\bibitem{efimov}
V.~Efimov, {\it Sov. J. Nucl. Phys.\/} {\bf 12}, 589 (1971). Originally: Yad.
  Fiz. \textbf{12}, 1080-1091 (1970).

\bibitem{innsbruck_efimov}
T.~Kraemer, {\it et~al.\/}, {\it Nature\/} {\bf 440}, 315 (2006).

\bibitem{innsbruck_atomdimer}
S.~Knoop, {\it et~al.\/}, {\it Nature Physics\/} {\bf 5}, 227 (2009).

\bibitem{wir_unten}
T.~B. Ottenstein, T.~Lompe, M.~Kohnen, A.~N. Wenz, S.~Jochim, {\it Phys. Rev.
  Lett.\/} {\bf 101}, 203202 (2008).

\bibitem{ohara_unten}
J.~H. Huckans, J.~R. Williams, E.~L. Hazlett, R.~W. Stites, K.~M. O'Hara, {\it
  Phys. Rev. Lett.\/} {\bf 102}, 165302 (2009).

\bibitem{lens_hetero}
G.~Barontini, {\it et~al.\/}, {\it Phys. Rev. Lett.\/} {\bf 103}, 043201
  (2009).

\bibitem{lens_homo}
M.~Zaccanti, {\it et~al.\/}, {\it Nature Physics\/} {\bf 5}, 586 (2009).

\bibitem{ohara_oben}
J.~R. Williams, {\it et~al.\/}, {\it Phys. Rev. Lett.\/} {\bf 103}, 130404
  (2009).

\bibitem{lev_li7}
N.~Gross, Z.~Shotan, S.~Kokkelmans, L.~Khaykovich, {\it Phys. Rev. Lett.\/}
  {\bf 103}, 163202 (2009).

\bibitem{randy_li7}
S.~E. Pollack, D.~Dries, R.~G. Hulet, {\it Science\/} {\bf 326}, 1683 (2009).

\bibitem{lev_li7_new}
N.~Gross, Z.~Shotan, S.~Kokkelmans, L.~Khaykovich, {\it Phys. Rev. Lett.\/}
  {\bf 105}, 103203 (2010).

\bibitem{wir_atomdimer}
T.~Lompe, {\it et~al.\/}, {\it Phys. Rev. Lett.\/} {\bf 105}, 103201 (2010).

\bibitem{naidon_atomdimer}
S.~Nakajima, M.~Horikoshi, T.~Mukaiyama, P.~Naidon, M.~Ueda, {\it Phys. Rev.
  Lett.\/} {\bf 105}, 023201 (2010).

\bibitem{greene_4body}
J.~von Stecher, J.~P. D'Incao, C.~H. Greene, {\it Nature Physics\/} {\bf 5},
  417 (2009).

\bibitem{innsbruck_4body}
F.~Ferlaino, {\it et~al.\/}, {\it Phys. Rev. Lett.\/} {\bf 102}, 140401 (2009).

\bibitem{bartenstein_julienne}
M.~Bartenstein, {\it et~al.\/}, {\it Phys. Rev. Lett.\/} {\bf 94}, 103201
  (2005).

\bibitem{braaten_6li}
E.~Braaten, H.~W. Hammer, D.~Kang, L.~Platter, {\it Phys. Rev. A\/} {\bf 81},
  013605 (2010).

\bibitem{jin_RF}
C.~A. Regal, C.~Ticknor, J.~L. Bohn, D.~S. Jin, {\it Nature\/} {\bf 424}, 47
  (2003).

\bibitem{arlt_association}
C.~Klempt, {\it et~al.\/}, {\it Phys. Rev. A\/} {\bf 78}, 061602 (2008).

\bibitem{pascal_french}
P.~Naidon, M.~Ueda, {\it http://arxiv.org/abs/1008.2260\/}  (2010).

\end{thebibliography}

\bibliographystyle{Science}


\begin{scilastnote}
\item We thank E. Braaten, J. P. D'Incao and H. W. Hammer for inspiring discussions. We thank P. Naidon for providing his data on the trimer binding energies.
  This work was supported by the Helmholtz Alliance HA216/EMMI and the Heidelberg Center for Quantum Dynamics. We are grateful to J. Ullrich and his group for their support. G. Z. and A. N. W. acknowledge support by the IMPRS-QD.\\
\end{scilastnote}

\section*{Supporting Online Material}

\section*{Materials and Methods}

We perform our measurements on $\left|2\right\rangle$-$\left|12\right\rangle$ and $\left|2\right\rangle$-$\left|23\right\rangle$ atom-dimer mixtures consisting of roughly $8.5\times 10^4$ atoms and $7.5\times 10^4$ dimers in an optical dipole trap with trapping frequencies  $( \omega_x, \omega_y,\omega_z)  \approx 2\pi \times (820,820,75)$\,Hz at a temperature of about $1\,\rm{\mu K}$, which we prepare as described in \cite{wir_unten, wir_atomdimer}. The homogenous magnetic field is created by a pair of Helmholtz coils. To associate trimers we apply RF-fields for 35-50 ms using an antenna resonant at $\sim 76$\,MHz driven by a 100\,W RF-amplifier. The Rabi frequency is $\Omega \approx 2 \pi \times 7$\,kHz ($~20$\,kHz) for the $\left|2\right\rangle$-$\left|3\right\rangle$ ($\left|1\right\rangle$-$\left|2\right\rangle$) transition. 
The decay rate of the trimer is constrained by a lower bound of $50$\,kHz derived from the width of the three-atom loss resonance at 895\,G\cite{braaten_6li}, however as the observed width of the atom-dimer resonance at 685\,G is larger the decay rate in the magnetic field region of interest is expected to be higher. From the width of the association features we obtain a decay rate of $~300$\,kHz, however this is only an upper bound, as the features are broadened by temperature and saturation effects.
Due to interference caused by the strong RF-fields, we have to deactivate the feedback of the magnetic field stabilization for the duration of the RF-pulse. This leads to a magnetic field uncertainty of at most $1$\,G. However, as the RF-transitions tune weakly with the magnetic field this only causes a shift of the bare transition which is small compared to the width of the features. After the RF-pulse we perform state-selective absorption imaging, which detects both free atoms in state $\left|2\right\rangle$ and - with slightly reduced efficiency - atoms in state $\left|2\right\rangle$ bound in $\left|12\right\rangle$ or $\left|23\right\rangle$ molecules.

When fitting the spectra obtained for the $\left|2\right\rangle$-$\left|12\right\rangle$ mixture we perform a least-squares fit of the data with a model of the form $N(\nu) = N_0 - A_0 e^{-\frac{(\nu-\nu_0)^2}{w_0^2}} - A_1 e^{-\frac{(\nu-\nu_1)^2  }{w_1^2}}$, where we exclude data points which are affected by dimer dissociation. The free parameters are  position, width and amplitude of the free-free ($\nu_0, w_0, A_0$) and association ($\nu_1, w_1, A_1$) peaks and the overall offset $N_0$. The given errors are the $2\sigma$ confidence bounds of the fit. However, due to the naive model for the lineshapes there can be additional systematic errors, which we estimate to be a fraction of the width $w_1 \approx 50$\, kHz of the association features. Temperature effects can cause an additional shift on the order of $\frac{3}{2}k_bT/h \approx 30$\,kHz. For the $\left|2\right\rangle$-$\left|23\right\rangle$ mixture we follow the same fitting procedure, except that we have to constrain the position of the free-free peak to the value calculated from the magnetic field to achieve a stable fit.
\clearpage

\begin{figure}
\centering
\includegraphics [width= 8cm] {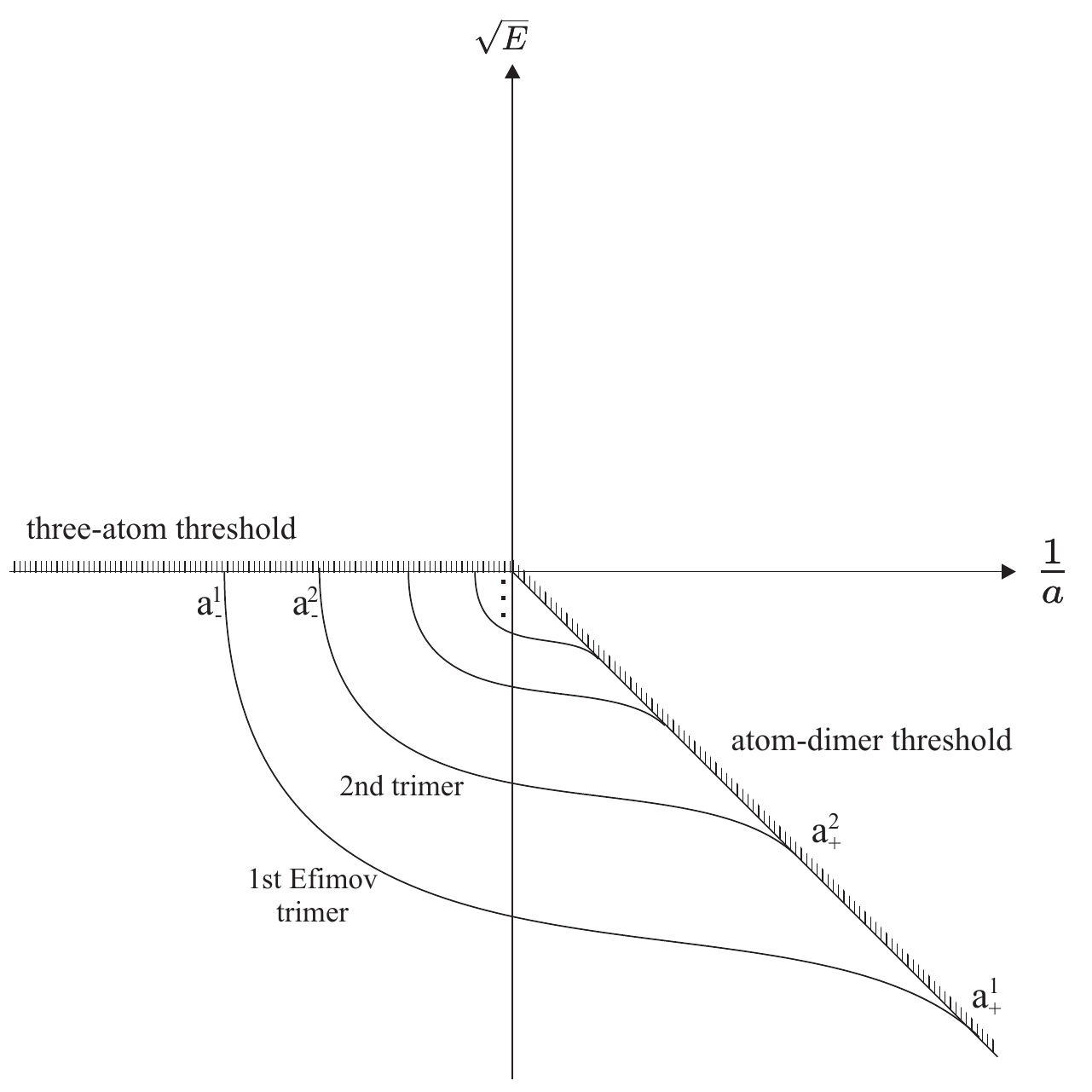}
\caption{Sketch of Efimov's scenario for the case of three identical scattering lengths. The square root of the binding energy of the universal dimer and trimer states is shown as a function of the inverse scattering length $1/a$. The crossings of the trimer states with the continuum follow a geometric scaling with a universal scaling factor $e^{\pi /s_0} \approx 22.7$, where $s_0 \approx 1.00624$ is a universal scaling parameter. For negative scattering length the first Efimov trimer crosses into the three-atom continuum at a critical scattering length $a_-^1$, followed by the second trimer state at $a_-^2 \approx 22.7 \, a_-^1$. If the scattering length diverges there is an infinite series of trimer states with exponentially decreasing binding energy. For positive scattering length these trimer states become unbound when they cross the atom-dimer threshold, i.e. their energy becomes degenerate with the energy of a dimer and a free atom. These crossings are spaced by the same universal scaling factor $e^{\pi /s_0}$.}
\label{fig:schema_efimov_scenario}
\end{figure}

\begin{figure}
\centering
\includegraphics [width= 8cm] {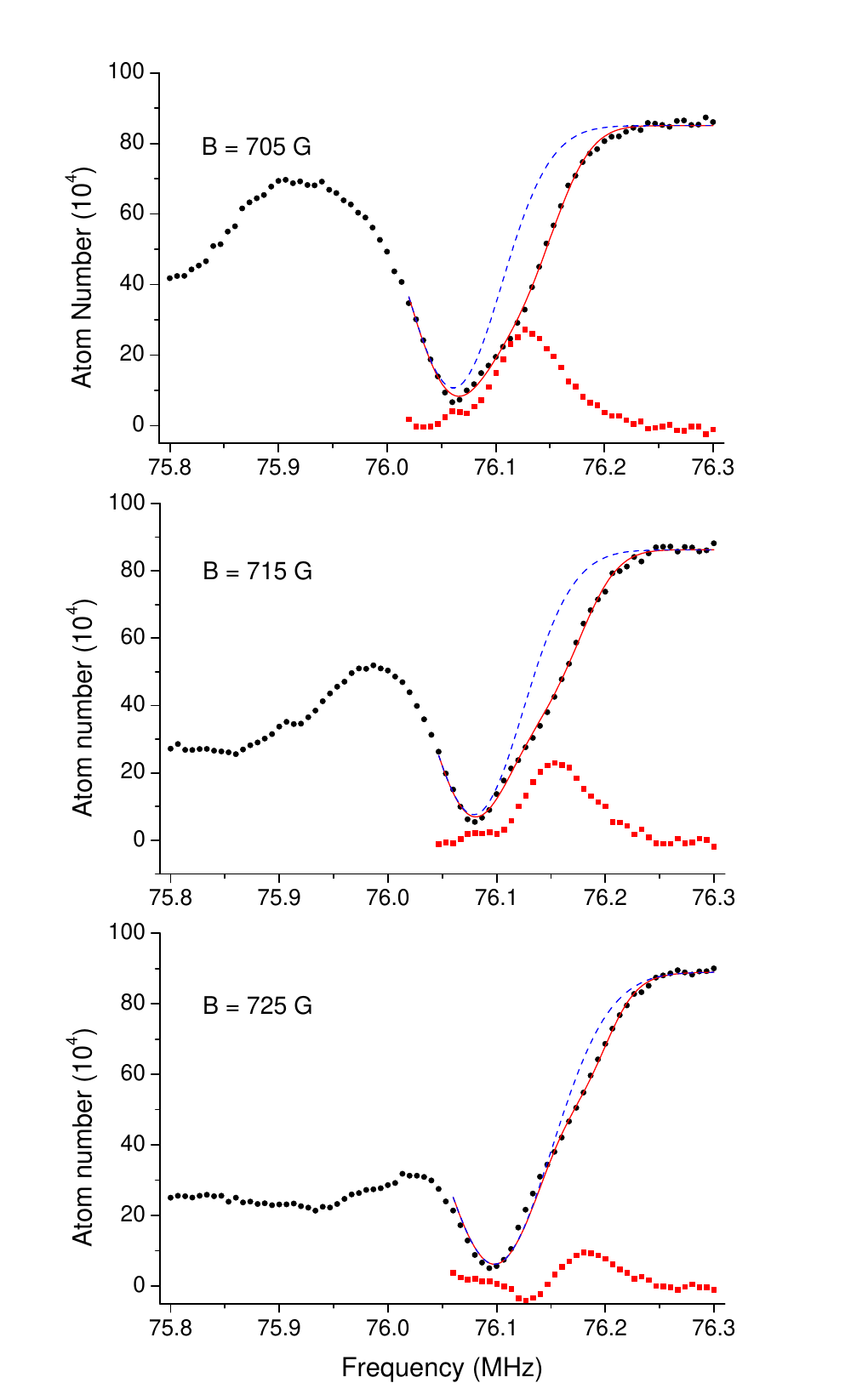}
\caption{Spectra for trimer association from a mixture of atoms in state $\left|2\right\rangle$ and $\left|23\right\rangle$ dimers (black circles).  The red lines are a fit of two overlapping Gaussians to the data. The fit to the central dip is shown as a blue line, the difference between this fit and the data is shown as red squares. As the free atoms are driven to state $\left|1\right\rangle$ instead of state $\left|3\right\rangle$ the features for dimer dissociation appear at lower frequency than the bare transition, while the trimer association appears for higher frequency. Due to the smaller separation between the bare transition and the trimer association the visibility of the association dip is not as good as for the $\left|2\right\rangle$-$\left|12\right\rangle$ mixture. For magnetic fields below 705\,G the association feature is too close the the bare transition to determine its position.  Each data point is the average of eight to nine individual measurements}
\label{fig:sample_plots_2_23}
\end{figure}

\end{document}